\documentclass[amsmath,amssymb,showkeys,showpacs]{revtex4}
\usepackage{amsmath}
\usepackage{graphicx}
\usepackage{rotating}%\usepackage{amsmath}
\def\dbox#1{\hbox{\vrule  %  Open box size 2#1 (Abrahams p 273) 
                        \vbox{\hrule \vskip #1
                             \hbox{\hskip #1
                                 \vbox{\hsize=#1}%
                              \hskip #1}%
                         \vskip #1 \hrule}%
                      \vrule}}
\def\qed{\hfill \dbox{0.05true in}}  %  QED
\begin{document}
%\hspace*{4 in}{[poly(2010).tex] $26^{\rm th}$ July 2010
\vspace*{0.4 in}
\title{Physical applications of second-order linear differential equations that admit polynomial solutions}
%\title{On polynomial solutions of a second-order linear differential equation and some applications}
\author{Hakan Ciftci$^{\ast }$, Richard L Hall$\dagger$, Nasser Saad$^\ddagger$ and Ebubekir Dogu}
\email{nsaad@upei.ca, hciftci@gazi.edu.tr, rhall@mathstat.concordia.ca}
\affiliation{$^{\ast }$Gazi \"{U}niversitesi, Fen-Edebiyat Fak\"{u}ltesi, Fizik B\"{o}l%
\"{u}m\"{u}, 06500 Teknikokullar-Ankara, Turkey.\\
$^\dagger$Department of Mathematics and Statistics, Concordia University,
1455 de Maisonneuve Boulevard West, Montr\'eal,
Qu\'ebec, Canada H3G 1M8\\
$^\ddagger$Department of Mathematics and Statistics,
University of Prince Edward Island,
550 University Avenue, Charlottetown,
PEI, Canada C1A 4P3.}%Lines break automatically or can be forced with \\

\begin{abstract}
\noindent Conditions for the second-order linear differential equation
$$\left(\sum_{i=0}^3 a_{3,i}x^i\right) y''+\left(\sum_{i=0}^2 a_{2,i}x^i\right) y'-\left(\sum_{i=0}^1 \tau_{1,i}x^i\right) y=0$$
to have polynomial solutions are given. Several application of these results to Schr\"odinger's equation are discussed. Conditions under which the confluent, biconfluent, and the general Heun equation yield polynomial solutions are explicitly given. Some new classes of exactly solvable differential equation are also discussed. The results of this work are expressed in such way as to allow direct use, without preliminary analysis.
\end{abstract}
\maketitle
\section{Introduction}

\noindent Since the early work of Bochner \cite{bochner} on the classification of polynomial solutions of second-order linear homogeneous differential equations, the problem of finding polynomial solutions to a given differential equation has attracted the attention of many researchers  \cite{bochner}-\cite{fryantsev}. A reason for such interest is that many problems in quantum mechanics, specially those arising from Schr\"odinger's equation, after the separation of the asymptotic-behavior factor of the wave function, there remains a polynomial-type factor in the solution \cite{chhajilany}- \cite{hall1}. Given a differential equation  
\begin{equation}\label{eq1}
{y}^{\prime \prime }=\lambda_0(x)y^\prime+s_0(x)y,
\end{equation}
it may appear to be a simple question to ask whether Eq.(\ref{eq1}) has a polynomial solution or not. In fact, unless the given differential equation lies within the classification scheme of Bochner  \cite{bochner}, the answer to this question seems far from simple. The problem seems even much harder if we ask whether Eq.(\ref{eq1}) has
polynomial solutions $P_n(x)$ not necessarily in a single variable $x$ but rather a polynomial-type solution in a {\it function} $f(x)$, say 
$$P_n(x)=\sum_{k=0}^n\alpha_k[f(x)]^k.$$
This problem was recently discussed by Hall et al in the context of finding exact solutions of Schr\"odinger's equation with soft-core Coulomb Potential \cite{hall1}. There are a number of well-known criteria \cite{bochner} for determining if there are polynomial solutions of the following differential equation 
\begin{equation}\label{eq2}
(a_{2,0}x^2+a_{2,1}x+a_{2,2}){y}^{\prime \prime}+(a_{1,0}x+a_{1,1})y'-\tau_{0,0}y=0,
\end{equation}
which we  now outline.
\vskip0.1true in
\noindent{\bf Theorem 1.} (\cite{saad}, Theorem 3) \emph{
The differential equation (\ref{eq2}) has a nontrivial polynomial solution of degree $n\in \mathbb N$ (the set of nonnegative integers) if, for fixed $n$,
\begin{equation}\label{eq3}
\tau_{0,0}=n(n-1)~a_{2,0}+n~a_{1,0},
\end{equation}
where the polynomials may be readily obtained from the recurrence relation ($n=0,1,2\dots$):
\begin{align}\label{eq4}
&y_{n+2}=\bigg[{((2n+1)a_{2,0}+a_{1,0})(2(n+1)a_{2,0}+a_{1,0})\over (na_{2,0}+a_{1,0})}x\notag\\
&+{((2n+1)a_{2,0}+a_{1,0})(2n(n+1)a_{2,0}a_{2,1}+2(n+1)a_{1,0}a_{2,1}-2a_{1,1}a_{2,0}+a_{1,0}a_{1,1})\over (na_{2,0}+a_{1,0})(2na_{2,0}+a_{1,0})}\bigg]y_{n+1}\notag\\
&+\left[{(n+1)(2(n+1)a_{2,0}+a_{1,0})(4n^2a_{2,2}a_{2,0}^2+a_{2,0}a_{1,1}^2+4na_{2,0}a_{1,0}a_{2,2}-n^2a_{2,0}a_{2,1}^2+a_{1,0}^2a_{2,2}-a_{1,1}a_{1,0}a_{2,1}-na_{1,0}a_{2,1}^2)\over (na_{2,0}+a_{1,0})(2na_{2,0}+a_{1,0})}\right]y_n.
\end{align} 
initiated with $$y_0=1,\quad y_1=a_{1,0}x+a_{1,1}.$$
}
\vskip 0.1true in
\noindent This theorem classifies most of the standard orthogonal polynomials such as Laguerre,
Hermite, Legendre, Jacobi, Chebyshev (first and second kind), Gegenbauer, and the
Hypergeometric type, etc. As a simple illustration of this Theorem, we consider the differential equation
\begin{equation*}
x^2{d^2y\over dx^2}+(2x+2){dy\over dx}-\tau_{0,0}y=0.
\end{equation*}
For polynomial solutions, we must have $\tau_{0,0}=n(n-1)+2n=n(n+1),\quad n\in\mathbb N$, and the recurrence relation generating the polynomial solutions now reads
\begin{equation*}
y_{n+2}=2(2n+3)xy_{n+1}+4y_n,\quad y_0=1,y_1=2x+2.
\end{equation*}
These polynomials were studied earlier by Krall and Frink and are known in the literature as Bessel polynomials  \cite{krall}.
\vskip0.1true in
\noindent{\bf Theorem 2:} (\cite{fryantsev}, Theorem 3) \emph{A polynomial $P_m(z)=\sum_{j=0}^m a_jz^j$ is a solution of the equation (\ref{eq2}) if and only if for any $z\in\mathbb C$ (the set of all complex numbers) and $n\geq m+2$ it satisfies the following mean-value formula:
\begin{equation}\label{eq5}
P_m(z)={2 \over 4nQ_2(z)+m(m-1){Q_2}''+2m{Q_1}'}\sum_{\mu=1}^n(Q_1(z)\lambda_\mu+2Q_2(z))P_m(z+\lambda_\mu)
\end{equation} 
where $Q_2(z)=a_{2,0}z^2+a_{2,1}z+a_{2,2}, Q_1(z)=a_{1,0}z+a_{1,1}$ and the numbers $\lambda_\mu = \lambda_\mu(n, 2)$ are roots of the equation  
\begin{equation}\label{eq6}
\sum_{\mu=0}^N (-1)^\mu {\lambda^{n-2\mu}\over 2^\mu \mu!}=0,\quad N=\left[{n\over 2}\right]
\end{equation}
for $m, k$, and $n$ be positive integers such that $n\geq k + m$ and the square brackets $\left[{\eta}\right]$ denote the integral part of a number $\eta$.
}
\vskip0.1true in
\noindent A polynomial $P_m(x)$ is a Bessel polynomial if and only if it satisfies the following mean-value
formula:
\begin{equation*}
P_m(x)=\sum_{\mu=1}^n{2((x+1)\lambda_\mu+x^2)\over 2nx^2+m(m+1)} P_m(x+\lambda_\mu),\quad n\geq m+2
\end{equation*}
where $\lambda_\mu$ are the roots of the equation
\begin{equation*}
\sum_{\mu=0}^N (-1)^\mu {\lambda^{n-2\mu}\over 2^\mu \mu!}=0,\quad N=\left[{n\over 2}\right].
\end{equation*}
\vskip0.1true in
\noindent In the present work, we find the conditions under which the differential equation 
\begin{equation}\label{eq7}
(a_{3,0}x^3+a_{3,1}x^2+a_{3,2}x+a_{3,3})~y^{\prime \prime}+(a_{2,0}x^2+a_{2,1}x+a_{2,2})~y'-(\tau_{1,0} x+\tau_{1,1})~y=0,
\end{equation}
has polynomial solutions of the form $y_n(x)=\sum\limits_{k=0}^n a_kx^k$. This class of equations contains a number of important differential equations such as the confluent  Heun equation, biconfluent  Heun equation and the general Heun equation \cite{heun}-\cite{fiziev}, which are usually studied individually in the literature. Some general results concerning solutions of Eq.(\ref{eq7}) may be found in Ref.\cite{pasquine}. Our method of studying the polynomial solutions of Eq.(\ref{eq7}) relies on the asymptotic iteration method (AIM), which can be summarized by means of the following two theorems.
\vskip0.1true in
\noindent{\bf Theorem 3:} (\cite{hakan}, equations (2.13)-(2.14)) \emph{Given $\lambda_0\equiv\lambda _{0}(x)$ and $s_0\equiv s_{0}(x)$ in $C^{\infty }$, the differential equation (\ref{eq1}) has the general solution%
\begin{equation}\label{eq8}
y=\exp \left( -\int\limits^{x}\alpha (t)dt\right) \left[ C_{2}+C_{1}\int%
\limits^{x}\exp \left( \int\limits^{t}\left( \lambda _{0}(\tau )+2\alpha
(\tau )\right) d\tau \right) dt\right]
\end{equation}%
if for some $n>0$ 
\begin{equation}\label{eq9}
\frac{s_{n}}{\lambda _{n}}=\frac{s_{n-1}}{\lambda _{n-1}}=\alpha (x)
,\quad\mbox{equivalently}\quad\delta _{n}(x)\equiv\lambda _{n}s_{n-1}-\lambda _{n-1}s_{n}=0,
\end{equation}%
where, for $n\geq 1$,
\begin{align}\label{eq10}
\lambda _{n}&=\lambda _{n-1}^{\prime }+s_{n-1}+\lambda_{0}\lambda _{n-1},\notag \\
 s_{n}&=s_{n-1}^{\prime }+s_{0}\lambda _{n-1}.
\end{align}
}
\vskip0.1true in
\noindent{\bf Theorem 4:} (\cite{saad}, Theorem 2) \emph{(i) If the second order differential equation (\ref{eq1}) has a polynomial solution of
degree $n$, then%
\begin{equation}\label{eq11}
\delta _{n}(x)=\lambda _{n}s_{n-1}-\lambda _{n-1}s_{n}=0,
\end{equation}%
where $\lambda_n$ and $s_n$ are given by (\ref{eq10}). Conversely (ii) if $\lambda _{n}\lambda _{n-1}\neq 0$ and $\delta _{n}(x)=0$%
, then the differential equation (\ref{eq1}) has the polynomial solution of degree at most $n$.
}
\vskip0.1true in
\noindent Theorem 4 may be the simplest criterion available in the literature for testing whether a given differential equation, such as (\ref{eq1}), has polynomial solutions or not. 
\vskip0.1true in
\noindent The conditions given in the present work are considered new as they generalize the earlier work of Krylov and Robnik  \cite{krylov}-\cite{Robnik} and they are explicitly written for practical use within the scope of physical applications. They answer directly, without further analysis, whether a given differential equation of the form (7) has polynomial solutions or not. Our results, especially theorem 5 below, are of important practical use in solving, for example, Schr\"odinger's equation where the construction of the eigenfunctions  for bound states, after extracting the asymptotic behaviors,  are based on the requirement of termination for some
hypergeometric series  \cite{krylov}-\cite{sever}.  Most of the available techniques are usually based on constructing classes of differential equations, see for example the construction of Turbiner \cite {turbiner} and the work of Krylov and Robnik \cite{Robnik}, that admit polynomial solutions. Very few results are available to test whether a given differential equation has actually a polynomial solution. The present results reported in the paper, Theorems 5 and 6, can be seen as criteria to test the solvability of a differential equation of the form (\ref{eq7}) in terms of polynomial solutions.
\vskip0.1true in

\noindent The present article is organized as follows. In the following section we find the conditions under which Eq.(\ref{eq7}) has polynomial solutions if there is any. Therein,  we discuss some examples found in mathematical physics literature, in order to demonstrate the validity of our general approach. In Section (III) we examine the polynomial solutions of the confluent, biconfluent and general Heun equation. In Section (IV) we discuss a special case of a more general  class of second-order differential equations 
  $$\left(\sum_{i=0}^{k+2} a_{k+2,i}x^i\right) y''+\left(\sum_{i=0}^{k+1} a_{k+1,i}x^i\right) y'-\left(\sum_{i=0}^k \tau_{k,i}x^i\right) y=0,\quad k=0,1,2,\dots$$
which admit polynomial solutions.
%%%%%%%%%%%%%%%%%%%%%%%%%%%%%%%%%%%%%%%%%%%%%%%%%%%%%%%%%%%%%%%%%
\section{Polynomial solutions of equation (\ref{eq7})}
%%%%%%%%%%%%%%%%%%%%%%%%%%%%%%%%%%%%%%%%%%%%%%%%%%%%%%%%%%%%%%%%%%%%
\noindent{\bf Theorem 5.} \emph{
The second-order linear differential equation (\ref{eq7}) has a polynomial solution of degree $n$ if, for any pair of the coefficients $a_{3,0}, a_{2,0}$ and $\tau_{1,0}$ not simultaneously zero,
\begin{equation}\label{eq12}
\tau_{1,0}=n(n-1)a_{3,0}+na_{2,0},\quad n=0,1,2,\dots,
\end{equation}
along with the vanishing of $(n+1)\times(n+1)$-determinant given by Table \ref{tab:table1}.
\begin{sidewaystable}
\centering
\caption{\label{tab:table1} The determinant $\Delta_{n+1}=0$ for the polynomial solutions of the differential equation (\ref{eq7}).\\}
\begin{tabular}{|llllllll|}
 $\tau_{1,1}~~$ & $-a_{2,2}~~$ & $-2a_{3,3}~~$ & 0~~&$\dots~~$&0~& 0~&0~ \\
  $\tau_{1,0}$ & $\tau_{1,1}-a_{2,1}$ &  $-2(a_{3,2}+a_{2,2})$& $-6a_{3,3}$&$\dots$&0&0&0 \\
0     & $\tau_{1,0}-a_{2,0}~~$  & $\tau_{1,1}-2(a_{3,1}+a_{2,1})~~$&$-3(2a_{3,2}+a_{2,2})~~$&$\dots$&0&0&0\\
$\vdots$&$\vdots$&$\vdots$&$\vdots$&$\vdots$&$\vdots$&$\vdots$&$\vdots$\\ 
0&0&0&0&$\dots$&$\tau_{1,0}-(n-2)(n-3)a_{3,0}-(n-2)a_{2,0}~~~$&$\tau_{1,1}-(n-1)(n-2)a_{3,1}-(n-1)a_{2,1}~~$&$-n((n-1)a_{3,2}+a_{2,2})$\\
0&0&0&0&$\dots$&0&$\tau_{1,0}-(n-1)(n-2)a_{3,0}-(n-1)a_{2,0}$&$\tau_{1,1}-n((n-1)a_{3,1}+a_{2,1})$\\
\end{tabular}
\end{sidewaystable}
}
\vskip0.1true in
\noindent{Proof:} The proof of this assertion follows by use of theorem 3, equation (\ref{eq9}), which yields for
\begin{align*}
\delta_1&=0\quad \mbox{if}\quad  \tau_{1,0}=a_{2,0}\quad \mbox{and}\quad\left| \begin{array}{ll}
  \tau_{1,1} & -a_{2,2} \\
\tau_{1,0} & \tau_{1,1}-a_{2,1}
  \end{array}\right| = 0\\ \\
\delta_2&=0\quad \mbox{if}\quad \tau_{1,0}=2a_{3,0}+2a_{2,0}\quad\mbox{and}\quad \left| \begin{array}{ccc}
  \tau_{1,1} & -a_{2,2} &  -2a_{3,3} \\
  \tau_{1,0} & \tau_{1,1}-a_{2,1} &  -2a_{3,2}-2a_{2,2} \\
  0    & \tau_{1,0}-a_{2,0}    & \tau_{1,1}-2(a_{3,1}+a_{2,1}) 
  \end{array}\right| = 0
\end{align*}
\begin{align*}
\delta_3&=0\quad\mbox{if}\quad \tau_{1,0}=6a_{3,0}+3a_{2,0}\quad\mbox{and}\quad \left| \begin{array}{cccc}
  \tau_{1,1} & -a_{2,2} &  -2a_{3,3} & 0 \\
  \tau_{1,0}&\tau_{1,1}-a_{2,1} &  -2a_{3,2}-2a_{2,2}& -6a_{3,3} \\
0     & \tau_{1,0}-a_{2,0}     & \tau_{1,1}-2a_{3,1}-2a_{2,1}&-6a_{3,2}-3a_{2,2}\\
0 &0& \tau_{1,0}-2a_{3,0}-2a_{2,0} & \tau_{1,1}-6a_{3,1}-3a_{2,1} 
  \end{array}\right| = 0
\end{align*}
and so on. A procedure which can be easily generalized for $\delta_n=0$ to yields $\tau_{1,0}=n(n-1)a_{3,0}+na_{2,0}$ subject to vanishing  $(n+1)\times(n+1)$-determinant $\Delta_{n+1}=0$ where  $\Delta_{n+1}$ given by Table \ref{tab:table1}. The derivation of this determinant can be obtained by substituting $f(x)=\sum_{k=0}^n c_k x^k$ to yield the four-term recurrence relation
\begin{align*}
\left[\tau_{1,0}-(k-2)(k-1)a_{3,0}-(k-1)a_{2,0}\right]c_{k-1}&+\left[\tau_{1,1}-k((k-1)a_{3,1}+a_{2,1})\right]c_{k}-(k+1)(ka_{3,2}+a_{2,2})c_{k+1}\\
&-(k+2)(k+1)a_{3,3}c_{k+2}=0,\quad c_{-1}=0
\end{align*}   
from which the determinant follows directly.
\qed
\vskip0.1true in

\noindent{\bf Theorem 6.} \emph{A necessary condition for the second-order linear differential equation 
\begin{align}\label{eq14}
\left(\sum_{i=0}^{k+2}a_{k+2,i}x^{k+2-i}\right)y''&+\left(\sum_{i=0}^{k+1}a_{k+1,i}x^{k+1-i}\right)y'-\left(\sum_{i=0}^{k}\tau_{k,i}x^{k-i}\right)y=0
\end{align}
to have a polynomial solution of degree $n$ is
\begin{align}\label{eq15}
\tau_{k,0}=n(n-1) a_{k+2,0}+na_{k+1,0},\quad k=0,1,2,\dots.
\end{align}
}

\noindent{Proof:} This result follows by putting $y(x)=P_n(x)$ where $P_n(x)=\sum\limits_{j=0}^n a_jx^j$ in (\ref{eq14}) and then multiply out everything and equating the coefficients of each power of  $x$. The highest power of $x$ yields the necessary condition (\ref{eq15}). \qed

\vskip0.1true in
\noindent{\bf Example 1:} (\cite{krylov}, equation 39) In their discussion of Schr\"odinger equations that allowing polynomial solutions,  Krylov and Robnik \cite{krylov} investigate the solution of the differential equation 
\begin{equation}\label{eq16}
x^3y''+\alpha(x^2-1)y+(\beta x+\gamma)y=0.
\end{equation}
The polynomial solutions of this equation follow directly by use of Theorem 5. Specifically, by using Eq.(\ref{eq12}), we find the following condition for polynomial solutions
\begin{equation}\label{eq17}
\beta=-n^2-(\alpha-1)n,\quad n=1,2,\dots
\end{equation}
where the parameters $\alpha$, $\beta$ and $\gamma$ must satisfy:
 
\begin{table}[th]
\centering
%\caption{\label{tab:table1} The determinant $\Delta_{N+1}=0$ for the polynomial solutions of Eq.(\ref{eq17}.}
\begin{tabular}{|lllllll|}
 $-\gamma~~$ & $\alpha$ & 0 & 0&\dots&0&0 \\
  $-\beta~~$ & $-\gamma$ &  $2\alpha~~$& 0&\dots&0&0 \\
0     & $-\beta-\alpha~~$     & -$\gamma$&$3\alpha$&\dots&0&0\\
\vdots&\vdots&\vdots&\vdots&\vdots&\vdots&\vdots\\ 
0&0&0&0&\dots~~&$-\beta-(n-1)(n-2)-(n-1)\alpha~~$&$-\gamma$ 
\end{tabular}~~=~0.
\end{table}

\noindent In particular, 
\begin{align*}
n&=1\Rightarrow \beta=-\alpha\quad\mbox{and}\quad\left| \begin{array}{ll}
  -\gamma & \alpha \\
\alpha & -\gamma
  \end{array}\right| = \gamma^2-\alpha^2=0\Rightarrow \gamma=\pm \alpha,\\
n&=2\Rightarrow \beta=-2(\alpha+1)\quad \mbox{and}\quad \quad \left| \begin{array}{ccc}
  -\gamma & \alpha & 0 \\
  -\beta & -\gamma &  2\alpha \\
  0    & -\beta-\alpha    & -\gamma 
  \end{array}\right| =-\gamma(2\alpha^2+3\alpha\beta+\gamma^2)= 0\\
~&\Rightarrow \gamma=0,\pm \sqrt{2\alpha(2\alpha+3)}.
\end{align*}
\vskip0.1true in 
\noindent{\bf Example 2:} (\cite{chhajilany}, Eq. (4)) In their study of the polynomial solutions of the planar Coulomb diamagnetic problem, Chhajlany and Malnev \cite{chhajilany} studied the following differential equation 
\begin{equation}\label{eq18}
y''+(p-2x^2)y'+(\delta x+\alpha)y=0.
\end{equation}
The equation admits polynomial solution of degree $n$ for 
\begin{equation}\label{eq19}
\delta=2n
\end{equation}
subject to the vanishing of the following determinant
\begin{equation}\label{eq20}
\left| \begin{array}{ccccccc}
  -\alpha & -p &  -2 & 0&\dots&0&0 \\
  -\delta & -\alpha &  -2p& -6&\dots&0&0 \\
0     & -\delta+2     & -\alpha&-3p&\dots&0&0\\
\vdots&\vdots&\vdots&\vdots&\vdots&\vdots&\vdots\\ 
0&0&0&0&\dots&-\delta+2(n-1)&-\alpha 
  \end{array}\right| = 0.
\end{equation}

\noindent{\bf Example 3:} (\cite{boztosun}, Eq. (49)) Boztosun \emph{et al.} studied analytical solutions of the Bohr Hamiltonian for the Davidson potential 
\begin{equation}\label{eq21}
V(\beta)=\beta^2+{\beta_0^2\over \beta^2}
\end{equation}
where $\beta_0$ is the position of the minimum of the potential (\cite{boztosun}, Eq. 3). Their analysis reduces to the investigation of the exact solutions of the differential equation (\cite{boztosun}, Eq. 49)
\begin{equation}\label{eq22}
f''_{n,L}(\beta)=(2\beta-{2(\mu+1)\over \beta})f'_{n,L}(\beta)+(2\mu+3-\epsilon)f_{n,L}(\beta),
\end{equation}
which can re-written,  by comparison with Eq. (\ref{eq7}), as
\begin{equation}\label{eq23}
\beta f''_{n,L}(\beta)-(2\beta^2-{2(\mu+1)})f'_{n,L}(\beta)-(2\mu+3-\epsilon)\beta f_{n,L}(\beta)=0.
\end{equation}
Using Theorem 5, for the polynomial solutions, we must have
$$\epsilon_n=2\mu+3+4n,\quad n=0,1,2,\dots.$$
The polynomial solutions using Eq.(\ref{eq8}) are 
$$
\left\{ \begin{array}{l}
y_0(x)= 1 \\
y_1(x)= 2x^2-3-2\mu\\
y_2(x)=4x^4-4(5+2\mu)x^2+(3+2\mu)(5+2\mu)\\
y_3(x)=8x^6-12(7+2\mu)x^4+6(7+2\mu)(5+2\mu)x^2-(3+2\mu)(5+2\mu)(7+2\mu)\\
\dots\\
       \end{array} \right.
$$

%%%%%%%%%%%%%%%%%%%%%%%%%%%%%%%%%%%%%%%%%%%%%%%%%%%%%%%%%%%%%%%%%%
\section{\protect\bigskip Polynomial solutions of the Heun Equation}
%%%%%%%%%%%%%%%%%%%%%%%%%%%%%%%%%%%%%%%%%%%%%%%%%%%%%%%%%%%%%%%%%%%
\subsection{Confluent Heun's Equation}
\noindent The confluent Heun's differential equation \cite{heun}-\cite{fiziev}, written in the simplest uniform shape, is
\begin{equation}\label{eq24}
y''+\left(\alpha+{\beta+1\over z}+{\gamma+1\over z-1}\right)y'+\left({\mu\over z}+{\nu\over z-1}\right)y=0
\end{equation}
which can be written as
\begin{equation}\label{eq25}
z(z-1)y''+\left(\alpha z^2+(\gamma+\beta-\alpha+2)z-\alpha+1\right)y'+\left((\mu+\nu)z-\mu\right)y=0.
\end{equation}
The polynomial solutions of this differential equation can be found easily using Theorem 5. For polynomial solutions, we must have
\begin{equation}\label{eq26}
\mu+\nu=-n\alpha,
\end{equation}
along with the vanishing of the $(n+1)\times(n+1)$-tridiagonal determinant given by

\begin{equation}\label{eq27}
\left| \begin{array}{ccccccc}
  \mu & \alpha-1 &  0 & 0&\dots&0&0 \\
  n\alpha & \mu-(\gamma+\beta-\alpha+2) &  2(\alpha-1)& 0&\dots&0&0 \\
0     & (n-1)\alpha     & \mu-2(\gamma+\beta-\alpha+3)&3(\alpha-1)&\dots&0&0\\
\vdots&\vdots&\vdots&\vdots&\vdots&\vdots&\vdots\\ 
0&0&0&0&\dots&\alpha&\mu-n(n-1+\gamma+\beta-\alpha+2) 
  \end{array}\right| = 0
\end{equation}

\subsection{Biconfluent Heun's equation}

\noindent For the biconfluent Heun's differential equation \cite{Ronveaux}
\begin{equation}\label{eq28}
xy''+\left(-2x^2-\beta x +(\alpha+1)\right)y'+\left((\gamma-\alpha-2)x-{1\over 2}(\delta+(\alpha+1)\beta)\right)y=0,
\end{equation}
the polynomial solutions can follow again by use of Theorem 5. The condition for polynomial solutions, in this case, is explicitly given by
\begin{equation}\label{eq29}
\gamma=\alpha+2(n+1),\quad n=0,1,2,\dots,
\end{equation}
along with the vanishing of the $(n+1)\times(n+1)$-tridiagonal determinant given by Table \ref{tab:table2}.

{\tiny
\begin{sidewaystable}
\centering
\caption{\label{tab:table2} The determinant $\Delta_{n+1}$ for the polynomial solutions of the differential equation (\ref{eq28}).\\}
\begin{tabular}{|llllllll|}
 ${1\over 2}(\delta+(\alpha+1)\beta)~~$ & $-(\alpha+1)~~$ & $0$ & 0~~&$\dots~~$&0~& 0~&0~ \\
~&~&~&~&$\dots$&~&$~$&$~$\\
  $-(\gamma-\alpha-2)$ & ${1\over 2}(\delta+(\alpha+1)\beta)+\beta~~$ &  $-2(\alpha+2)$& $0$&$\dots$&0&0&0 \\
~&~&~&~&$~$&~&$~$&$~$\\
0     & $-(\gamma-\alpha-2)+2~~$  & ${1\over 2}(\delta+(\alpha+1)\beta)+2\beta~~$&$-3(\alpha+3)~~$&$\dots$&0&0&0\\
~&~&~&~&$~$&~&$~$&$~$\\
$\vdots$&$\vdots$&$\vdots$&$\vdots$&$\vdots$&$\vdots$&$\vdots$&$\vdots$\\ 
~&~&~&~&$~$&~&$~$&$~$\\
0&0&0&0&$\dots$&$-(\gamma-\alpha-2)~~~$&${1\over 2}(\delta+(\alpha+1)\beta)~~$&$-n(\alpha+n)$\\
~&~&~&~&~&$+2(n-2)$&$-(n-1)(n-2)+(n-1)\beta$& \\
~&~&~&~&$~$&~&$~$&$~$\\
0&0&0&0&$\dots$&0&$-(\gamma-\alpha-2)$&${1\over 2}(\delta+(\alpha+1)\beta)\beta$\\
~&~&~&~&$~$&0&$+2(n-1)$&$-n(n-1)+n\beta$\\
\end{tabular}
\end{sidewaystable}
}

\subsection{Heun's General Equation}
\noindent The Heun differential equation \cite{heun}-\cite{fiziev} with four regular singular points located at $z=0,1,a,\infty$  is given by
\begin{equation}\label{eq29}
y''+\left({\gamma\over z}+{\delta\over z-1}+{\epsilon\over x-a}\right)y'+{(\alpha\beta z-q)\over z(z-1)(z-a)}y=0,
\end{equation}
where the parameters satisfy the Fuchsian condition 
\begin{equation}\label{eq30}
1+\alpha+\beta=\gamma+\delta+\epsilon
\end{equation}
 to ensure the regularity of the singular point at infinity. Equation (\ref{eq29}) can be written as
\begin{equation}\label{eq31}
(z^3-(1+a)z^2+az)y''+\left((\gamma+\epsilon+\delta) z^2-(a(\delta+\gamma)+\epsilon+\gamma)z+a\gamma\right)y'+(\alpha\beta z-q)y=0
\end{equation}
The polynomial solutions of this differential equation follow by use of Theorem 5. For polynomial solutions, we must have
\begin{equation}\label{eq32}
\alpha\beta=-n(n-1)-n(\gamma+\epsilon+\delta)\overset{Eq.(\ref{eq31})}{\Rightarrow} \alpha=-n\quad\mbox{or}\quad \beta=-n,
\end{equation}
along with the vanishing of the $(n+1)\times(n+1)$-tridiagonal determinant given explicitly in Table (\ref{tab:table3}).

\begin{sidewaystable}
\centering
\caption{\label{tab:table3} The determinant $\Delta_{n+1}$ for the polynomial solutions of Heun's equation (\ref{eq30}).\\}
\begin{tabular}{|lllllll|}
 $q$ & -$a\gamma$ & 0 & 0&$\dots$&0&0 \\
  -$\alpha\beta$ & $q+(a(\delta+\gamma)+\epsilon+\gamma)$ &  -$2a(1+\gamma)$& 0&$\dots$&0&0 \\
0     & -$\alpha\beta-(\gamma+\epsilon+\delta)$     & $q+2(a+1)+2(a(\delta+\gamma)+\epsilon+\gamma))$&$-3(-2(a(\delta+\gamma)+\epsilon+\gamma)+a\gamma)$&$\dots$&0&0\\
$\vdots$&$\vdots$&$\vdots$&$\vdots$&$\vdots$&$\vdots$&$\vdots$\\ 
0&0&0&0&$\dots$&$-\alpha\beta-(n-1)[(n-2)$&$q+n(n-1)(1+a)$\\
~&~&~&~&~&$+(\gamma+\epsilon+\delta)]$&$+n(a(\delta+\gamma)+\epsilon+\delta)$ \\
\end{tabular}~~=~0
\end{sidewaystable}
\subsection{Shifted Coulomb Potential}

\noindent As an example of a Heun equation that can be treated directly using our results of Theorem 5, we consider the Schr\"odinger equation for a single particle bound by the shifted Coulomb potential
\begin{equation}\label{H1}
\left[-{1\over 2}\Delta+V_1(r)\right]\psi(r)=E\psi(r),\quad V_1(r)=-{Z\over r+\beta},\quad \beta>0
\end{equation}
where $\Delta$ is the $d-$dimensional Laplacian operator, $d\ge 2$. The potential $V_1(r)$ has been used as an approximation for
the potential due to a smeared charge distribution, rather than a point charge, and may be appropriate for describing
mesonic atoms \cite{ray}. This problem has been discussed  at length for $d=3$ in the two recent articles \cite{saad1}-\cite{saad2}. The purpose of this section is first to show that the quasi-exact solutions of the eigen-equation (\ref{H1}) follows directly from Theorem 5, and also to extend the earlier results of Hall et al.  \cite{saad1}-\cite{saad2} to arbitrary $d>1$ dimensions. In atomic units, the radial Schr\"odinger equation (\ref{H1}) for the potential $V_1(r)$ in $d$ dimensions reads
%\begin{equation}\label{H2}
%\left[-{1\over 2}{d^2\over dr^2}+{(2l+d-1)(2l+d-3)\over 8r^2}-{Z\over r+\beta}\right]\psi=E\psi
%\end{equation}
\begin{equation}\label{H2}
\left[-{1\over 2}{d^2\over dr^2}+{k(k+1)\over 2r^2}-{Z\over r+\beta}\right]\psi=E\psi,\quad k= {1\over 2}(2l+d-3)
\end{equation}
We may assume the solution of equation (\ref{H2}), which vanishes at the origin and at infinity, is
\begin{equation}\label{H3}
\psi(r)=r^{k+1}e^{-\alpha (r+\beta)}f(r+\beta),
\end{equation}
where $f(r+\beta)$ is to be determined.  On substituting Eq.(\ref{H3}) into Eq.(\ref{H2}), we obtain the following second-order differential equation for $f\equiv f(r+\beta)$
\begin{equation}\label{H4}
f''+2\left[{(k+1)\over r}-\alpha\right]f'+\left[-{2\alpha(k+1)\over r}+\alpha^2+{2Z\over r-\beta}+2E\right]f=0.
\end{equation}
We multiply by $r(r-\beta)$ to obtain for $E=-\alpha^2/2$
\begin{equation}\label{H5}
r(r+\beta)f''+\left[-2 \alpha r^2+2(k+1-\alpha\beta) r+2 \beta (1+k)\right]f'+\left[(-2\alpha (k+1)+2 Z)r-2 \alpha \beta (k+1)\right]f=0.
\end{equation}
Eq.(\ref{H5}) is an example of confluent Heun equation \cite{saad1}, however, theorem 5 gives directly the conditions for polynomial-type solutions, namely, 
\begin{equation}\label{H6}
\alpha={Z\over n'+k+1},\quad n'=n=0,1,2,\dots\quad {fixed}
\end{equation}
subject to the conditions on the potential parameters given by the vanishes of the tridiagonal determinant
\begin{center}
$\Delta_{n+1}$~~=~~\begin{tabular}{|lllllll|}
 $\beta_0~~$ & $\alpha_1$ &~&~& ~&~ &~\\
  $\gamma_1$ & $\beta_1$ &  $\alpha_2$&$~$&~&~&~ \\
~ & $\gamma_2$  & $\beta_2$&$\alpha_3$&~&~&~\\
$~$&~&$\ddots$&$\ddots$&$\ddots$&$~$&~\\ 
~&~&~&$\gamma_{n-2}$&$\beta_{n-2}$&$\alpha_{n-1}$&$~$\\
~&~&~&&~$\gamma_{n-1}$&$\beta_{n-1}$&$\alpha_n$\\
~&~&~&~&$~$&$\gamma_{n}$&$\beta_n$\\
\end{tabular}~~=~~0
\end{center}
where
\begin{align}\label{H7}
\beta_n&=2\alpha\beta(k+n+1)-n(n+2k+1)\notag\\
\alpha_n&=-n\beta(n+2k+1)\notag\\
\gamma_n&=2\alpha(n-n'-1)
\end{align}
The meaning of $n'$ is that $n'=n$ but fixed in the sense that for $\Delta_3=0$, we should fix $n'=2$ for our computations. Equation (\ref{H6}) gives the exact eigenvalues 
\begin{equation}\label{H8}
E_{nl}^d=-{1\over 2}{Z^2\over (n+k+1)^2},\quad n=0,1,2,\dots, k= {1\over 2}(2l+d-3), d\geq 2
\end{equation}
subject to conditions on the parameters $\alpha,\beta$ and $Z$ as given by the vanishing of the determinant $\Delta_{n+1}=0$. 
In Table \ref{tab:table4}, we summarize the first few conditions, for $n=1,2,3,4$, on the parameter $\beta$ in the potential $V_1(r)=Z/(r+\beta)$ so that Eq. (\ref{H5}) has polynomial solution of the form $f_n(r)$. It should be noted that theorem 5 tells us explicitly whether or not the differential equation has polynomial solutions: finding these polynomials is a problem that in this case remains to be solved, for example by using AIM.
 
\begin{table}
\caption{\label{tab:table4} Conditions on $\beta$ for the existence of polynomial solutions of Eq.(\ref{H5}).}
%\begin{ruledtabular}
\begin{tabular}{l|l} \hline
 $\alpha$& Conditions on $\beta$\\
\hline
$\frac{Z}{k+2}$& $\alpha\beta-1=0$ \\
\\
$\frac{Z}{k+3}$& $(k+2)\alpha^2\beta^2-3(k+2)\alpha\beta+2k+3=0$ \\
\\
$\frac{Z}{k+4}$& $(k+3)(k+2)\alpha^3\beta^3-6(k+3)(k+2)\alpha^2\beta^2+(11k^2+50k+54)\alpha\beta-3(k+2)(2k+3)=0$ \\
\\
$\frac{Z}{k+5}$& $(k+2)(k+3)(k+4)\alpha^4\beta^4-10(k+2)(k+3)(k+4)\alpha^3\beta^3+(720+823k+300k^2+35k^3)\alpha^2\beta^2$\\
&$-(720+925k+381k^2+50k^3)\alpha\beta+6(k+2)(2k+3)(2k+5)=0$\\ \hline
\end{tabular}
%\end{ruledtabular}
\end{table}

%%%%%%%%%%%%%%%%%%%%%%%%%%%%%%%%%%%%%%%%%%%%%%%%%%%%%%%%%%%%%%%%
\section{\protect\bigskip New classes of polynomial solutions}
%%%%%%%%%%%%%%%%%%%%%%%%%%%%%%%%%%%%%%%%%%%%%%%%%%%%%%%%%%%%%%%%
\noindent The differential equations discussed in the previous sections were characterized by the fact that the parameters $\tau_{1,0}$ and $\tau_{1,1}$ are the only coefficients in the differential equation that depend on the nonnegative integer $n$. The parameters $a_{3,i},i=0,1,2,3$ and $a_{2,j},j=0,1,2$ are constants to be determined, based on the values of $\tau_{1,k},k=0,1$ allowing polynomial solutions. In this section we use AIM to discuss new classes of differential equation, where the parameters $a_{2,i},i=0,1,2$ are also functions of $n$ that allow polynomial solutions.  We have 
\vskip0.1true in
\noindent\textbf{Theorem 7:} \emph{For arbitrary values $a$ and $b$ and nonnegative integers of $m$ and $n$, the  solutions of the second-order linear differential equation
\begin{equation}\label{eq33}
y''={ax^{l-2}\over b+{ax^{l-1}\over m+n}} y'+ {m(m+1)b\over x^2(b+{ax^{l-1}\over m+n})} y,\quad\quad l=2,3,\dots
\end{equation}
are given by
\begin{equation}\label{eq34}
y_n^{m}=x^{m+1\over l-1}{}_2F_1(-{n\over l-1},{m+1\over l-1};{2m+l\over l-1};-{ax^{l-1}\over b(n+m)}),~~m=1,2,3,\dots, n=0,1,2,\dots,
\end{equation} 
where, for polynomial solutions, $n=\nu(l-1)$, for some $\nu=0,1,2,\dots$. Here $_2F_1(\alpha,\beta;\gamma;x)$ refer to the classical Gauss hypergeometric series:
\begin{equation}\label{eq35}
{}_2F_1(\alpha,\beta;\gamma;x)=1+\sum_{k=1}^\infty {\alpha(\alpha+1)\dots(\alpha+k-1)~\beta(\beta+1)\dots(\beta+k-1)\over \gamma(\gamma+1)\dots(\gamma +k-1)}{x^k\over k!}.
\end{equation} 
In particular, for the differential equation ($l=2$):
\begin{equation}\label{eq36}
y''={a\over b+{ax\over m+n}} y'+ {m(m+1) b\over x^2(b+{ax\over m+n})} y,
\end{equation}
the polynomial solutions are given by
\begin{equation}\label{eq37}
y_n^{m}=x^{m+1}{}_2F_1(-n,m+1;2(m+1);-{{ax\over b(m+n)}}),~~ m=1,2,3,\dots, n=0,1,2,\dots.
\end{equation} 
Here, each $y_n^m$ is a polynomial of degree $n$ in the variable $x$.
}
\vskip0.1true in
\noindent{Proof:} By means of the substitution $z=b+{ax^{l-1}\over m+n}$, the differential equation Eq.(\ref{eq33}) is reduced to  
\begin{equation}\label{eq38}
{d^2y\over dz^2}=\left[{(n+m)\over (l-1)z}-{(l-2)\over (l-1)(z-b)}\right]{dy\over dz}+ {m(m+1)b\over (l-1)^2z(z-b)^2} y.
\end{equation}
Further, let 
$$y_n^m=(z-b)^{m+1\over l-1}f_n^m(z),$$
we find that the functions $f_n^m(z)$ now satisfy the differential equation
\begin{equation}\label{eq39}
f''(z)={\left[{(n-m-l)}z-{(n+m)b}\right]\over (l-1)z(z-b)}f'(z)+{n(m+1)\over (l-1)^2z(z-b)}f(z),
\end{equation}
which can easily be solved by use of the asymptotic iteration method, Theorem 3.
% --------------------------------------------
\section{Conclusion}
% --------------------------------------------
This paper is about exact solutions of second-order linear homogeneous differential equations that arise in mathematical physics. 
If the exact solutions are transcendental functions, then the term `exact' has a recursive ring to  it,
since such solutions can only be regarded as exact if the solution functions turn out to have been already named and studied.
An exception occurs  when the unknown part of a solution is a factor that is a polynomial, either
in the independent variable, say $x$, or in a function of it, $f(x).$  In the study of physical problems where the theory leads
to a differential equation, such as Schr\"odinger's equation, it is extremely  helpful to have at one's disposal some compactly-expressed solutions that typify the kind of exact system behaviour that is implied by the differential equation.  It is expected that the analysis presented in this paper will provide for a rich and useful variety of such exact solutions.

% ------------------------------------------------------
\section*{Acknowledgments}
% ------------------------------------------------------
\medskip
\noindent Partial financial support of this work under Grant Nos. GP3438 and GP249507 from the
Natural Sciences and Engineering Research Council of Canada is gratefully
acknowledged by two of us (respectively [RLH] and [NS]). 

%%%%%%%%%%%%%%%%%%%%%%%%%%%%%%%%%%%%%%%%%%%%%%%%%%%%%%%%%%%%%%%%%%%%%%%%%%%%%%%%%%%%%%%%%%%%%%%%%%%%%%%%%%%

\end{document}